\documentclass[pra,letterpaper,twocolumn,aps,footinbib,superscriptaddress,showpacs,floatfix]{revtex4-1}

\usepackage{amsmath, upgreek, amssymb, graphicx, subfigure, paralist, dsfont}
\usepackage[colorlinks, linkcolor=blue, citecolor=blue, urlcolor=blue, breaklinks=true]{hyperref}

\usepackage{tikz}
\usetikzlibrary{matrix,arrows}

\newcommand{\eref}[1]{Eq.~(\ref{#1})}
\newcommand{\erefs}[1]{Eqs.~(\ref{#1})}
\newcommand{\fref}[1]{Fig.~\ref{#1}}

\newcommand{\ie}{\emph{i.e.}}
\newcommand{\eg}{\emph{e.g.}}

\newcommand{\rmd}{\text{d}}

\newcommand{\define}{=}
\newcommand{\abs}[1]{\lvert#1\rvert}
\newcommand{\expt}[1]{\langle#1\rangle}

\newcommand{\im}[1]{\,\text{Im}\!\left\{#1\right\}}
\newcommand{\re}[1]{\,\text{Re}\!\left\{#1\right\}}

\makeatletter
\newlength \figurewidth
\makeatother

\begin{document}

\title{Dissipative Optomechanics in a Michelson--Sagnac Interferometer}
\author{Andr\'e Xuereb}
\email[Corresponding author.\ ]{andre.xuereb@aei.mpg.de}
\affiliation{Institut f\"ur Gravitationsphysik, Leibniz Universit\"at Hannover, Callinstra\ss{}e 38, D-30167 Hannover, Germany}
\affiliation{Institut f\"ur Theoretische Physik, Leibniz Universit\"at Hannover, Appelstra\ss{}e 2, D-30167 Hannover, Germany}
\author{Roman Schnabel}
\affiliation{Institut f\"ur Gravitationsphysik, Leibniz Universit\"at Hannover, Callinstra\ss{}e 38, D-30167 Hannover, Germany}
\author{Klemens Hammerer}
\affiliation{Institut f\"ur Gravitationsphysik, Leibniz Universit\"at Hannover, Callinstra\ss{}e 38, D-30167 Hannover, Germany}
\affiliation{Institut f\"ur Theoretische Physik, Leibniz Universit\"at Hannover, Appelstra\ss{}e 2, D-30167 Hannover, Germany}

\date{\today}


\begin{abstract}
Dissipative optomechanics studies the coupling of the motion of an optical element to the \emph{decay rate} of a cavity. We propose and theoretically explore a realization of this system in the optical domain, using a combined Michelson--Sagnac interferometer, which enables a strong \emph{and tunable} dissipative coupling. Quantum interference in such a setup results in the suppression of the lower motional sideband, leading to strongly enhanced cooling in the non-sideband-resolved regime. With state-of-the-art parameters, ground-state cooling and low-power quantum-limited position transduction are both possible. The possibility of a strong and tunable dissipative coupling opens up a new route towards observation of fundamental optomechanical effects such as ponderomotive squeezing or nonlinear dynamics. Beyond optomechanics, the method suggested here can be readily transferred to other setups involving such systems as nonlinear media, atomic ensembles, or single atoms.
\end{abstract}

\maketitle

Recent progress in the engineering of high-quality micromechanical oscillators coupled to high-finesse cavity modes has paved the way towards sensing and control of mechanical motion at the quantum limit~\cite{Kippenberg2008,Marquardt2009,Genes2009d,Aspelmeyer2010,Regal2011}. The rapid developments in the field of optomechanics bear important implications for both applied and basic science, ranging from applications in high-sensitivity metrology~\cite{Regal2008,Teufel2009,Anetsberger2010} and quantum information processing~\cite{Chang2009b,Hammerer2009b,Stannigel2010,Mazzola2011,Chang2011} to fundamental tests of quantum mechanics at large mass- and length-scales~\cite{Corbitt2007,Abbott2009,RomeroIsart2011b}.
\par
In the conventional paradigm of optomechanics the interaction of the mechanical oscillator with a cavity mode is \emph{dispersive} in the sense that the cavity resonance frequency experiences a shift depending on the displacement of the mechanical oscillator arising from conservative radiation pressure or optical gradient forces. This coherent dispersive interaction has been employed for sideband-cooling to the quantum mechanical ground state~\cite{Chan2011,Teufel2011}, as well as for the observation of optomechanical normal-mode splitting~\cite{Dobrindt2008,Groblacher2009a,Teufel2011b} and optomechanically-induced transparency~\cite{Weis2010,SafaviNaeini2011}. The complementary paradigm of \emph{dissipative} coupling---where the \emph{width} $\kappa_\text{c}$ of the cavity resonance, rather than its frequency $\omega_\text{c}$, is dependent on the mechanical displacement $x$---was introduced very recently in a theoretical study~\cite{Elste2009} in the context of electromechanics. This situation, rather unusual for cavity quantum electrodynamics, was shown~\cite{Elste2009} to give rise to remarkable quantum noise interference effects which dramatically relax requirements for cooling to the ground state without sideband resolution; it also allows reaching the standard quantum limit (SQL) for the imprecision in position measurements. While it is already clear that a strong and tunable dissipative optomechanical coupling would greatly enrich the toolbox of optomechanics, its full significance for the quantum control of optomechanical systems, \eg, for ponderomotive squeezing~\cite{Brooks2011}, nonlinear dymanics~\cite{Nunnenkamp2011,Rabl2011}, or pulsed protocols~\cite{Vanner2010}, is yet to be explored.\\
Unfortunately, an optomechanical setup having a strong dissipative coupling in the absence of dispersive coupling has not been found. Consider, \eg, a Fabry--P\'erot interferometer (FPI) of length $L$ and resonance frequency $\omega_\text{c}=\pi nc/L$ ($n\in\mathds{N}$), with one movable ideal end-mirror and an input coupler of transmissivity $\tau$, such that the cavity linewidth is $\kappa_\text{c}=c\abs{\tau}^2/(4L)$. When the mirror moves, the cavity length changes, and with it both $\omega_\text{c}$ and $\kappa_\text{c}$. The corresponding shifts per zero-point fluctuation $x_0$ of the mirror oscillator, $g_\omega=(\partial\omega_\text{c}/\partial x)x_0$ and $g_\kappa=(\partial\kappa_\text{c}/\partial x)x_0$, quantify the strength of dispersive and dissipative couplings, respectively. Their ratio is thus given by the cavity's quality factor, $g_\omega/g_\kappa=\omega_\text{c}/\kappa_\text{c}\gg 1$, such that the dispersive coupling dominates by far. A similar conclusion applies for a movable membrane coupled to a single mode of a FPI. One may obtain $g_\kappa\simeq g_\omega$ by coupling the membrane to multiple~\cite{Thompson2008,Sankey2010} transverse modes, such that \emph{both} types of coupling contribute to the dynamics; however in such a setup, one cannot `switch off' the dispersive interaction to take advantage of the quantum noise interference effects present in a purely dissipative coupling.
\begin{figure}[t]
 \centering
 \includegraphics[width=\figurewidth]{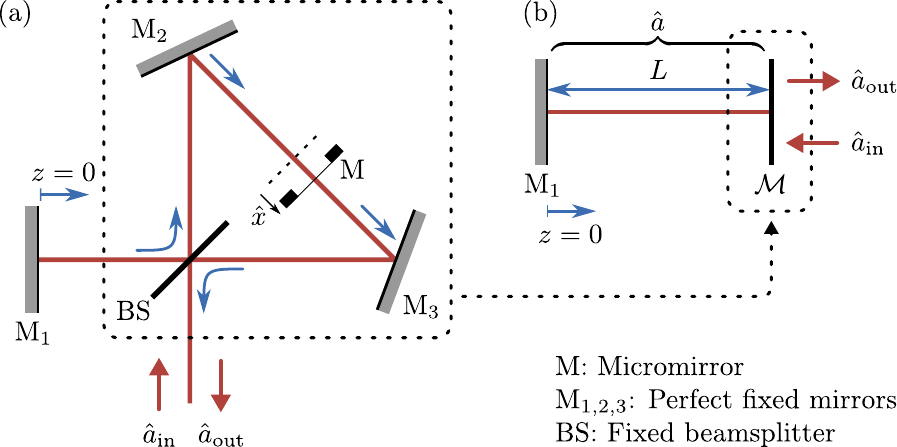}
\caption{(a)~Topology of the Michelson--Sagnac interferometer. (b)~Effective cavity; the properties of $\mathcal{M}$ depend on $\mathrm{M}$.}
 \label{fig:PhysicalModel}
\end{figure}
\par
Here we show that in a Michelson-Sagnac interferometer (MSI) with a movable membrane~\cite{Yamamoto2010,Friedrich2011}, cf.\ \fref{fig:PhysicalModel}, a strong and tunable optomechanical coupling can be achieved for which $g_\omega=0$ but where the effective \emph{dissipative} coupling strength~\cite{Groblacher2009a} can be of the order of the cavity linewidth. The idea is as follows: Consider the MSI operated at a point where the transmissivity $\tau$ of the effective mirror is close to zero; $\tau$ will then depend sensitively on the membrane displacement $x$. Combining this compound `MSI-mirror' with a perfect mirror in its dark port will result in an effective FPI whose linewidth depends on $x$ dominantly via $\tau$ and not via the change in the (effective) cavity length $L$, see~\fref{fig:PhysicalModel}. In contrast to a true FPI, therefore, $g_\kappa$ and $g_\omega$ have a different functional dependence on $x$. This feature gives rise to a topology (i)~where the optomechanical interaction can be \emph{tuned at will} between being strongly dissipative or dispersive, and (ii)~which can realize dissipative optomechanics leading to \emph{ground-state cooling} of the mechanical oscillator via quantum noise interference, as discussed in Ref.~\cite{Elste2009}. The distinctive signature of this interaction, an asymmetric Fano lineshape, is observable in the spectrum of the cavity output field. We also report on the suitability of using this system for (iii)~\emph{sensitive position transduction}.
\par
This paper is structured as follows. We shall first describe the physical model and write down the resulting Hamiltonian and input--output relation. These are then used to derive the equations of motion for the cavity field and oscillator motion. The resulting dynamics is solved to obtain the steady-state mechanical occupation number analytically in the weak-coupling limit, and also numerically in both weak- and strong-coupling regimes.

\section{Model}
The Hamiltonian including both dispersive and dissipative effects in an opto- or electromechanical~\cite{Elste2009} system can be written~(see Appendix)
\begin{align}
\label{eq:HOM}
\hat{H}_\text{OM}=\ & \bigl(-\Delta+g_\omega\hat{x}\bigr)\hat{a}^\dagger\hat{a}+\omega_\text{m} b^\dagger b+\int\!\!\rmd\omega\,\omega\,a^\dagger_\omega a_\omega\\
&+i\left(\sqrt{2\kappa_\text{c}}
+\frac{g_\kappa}{\sqrt{2\kappa_\text{c}}}\,\hat{x}\right)\int\!\!
\frac{\rmd\omega}{\sqrt{2\pi}}\bigl(\hat{a}^\dagger_\omega\hat{a}-\mathrm{H.c.}\bigr)\,.\nonumber
\end{align}
This Hamiltonian is written in a frame rotating at the optical frequency of some driving field $\omega_\text{d}$; $\Delta=\omega_\text{d}-\omega_\text{c}$ is the detuning from cavity resonance ($\omega_\text{c}$) whose annihilation operator is $\hat{a}$. $\hat{x}=(\hat{b}+\hat{b}^\dagger)/\sqrt{2}$ is the dimensionless displacement, and $\hat{b}$ is the annihilation operator for the mechanical oscillator of frequency $\omega_\text{m}$. The second line in \eref{eq:HOM} provides the coupling of the cavity mode $\hat{a}$ to the modes $\hat{a}_\omega$ of the external field coupling into and out of the effective FPI, and will give rise to the finite width $\kappa_\text{c}$ of the resonance. The two terms proportional to $\hat{x}$ describe the shifts of the cavity resonance and width with mechanical displacement, with strengths characterized by $g_{\omega,\kappa}$, respectively. This Hamiltonian generalizes the dispersive-only Hamiltonian that is considered in most works on optomechanical systems. Before we proceed to make the connection of Hamiltonian~\eqref{eq:HOM} and the physics of a MSI~\cite{Yamamoto2010,Friedrich2011} we note that the standard cavity-input--output relation must be generalized to accommodate the effect of dissipative coupling~(see Appendix):
\begin{equation}
\hat{a}_\text{out}-\hat{a}_\text{in}=\bigl[\sqrt{2\kappa_\text{c}}+\bigl(g_\kappa/\sqrt{2\kappa_\text{c}}\bigr)\hat{x}\bigr]\hat{a}\,.
\end{equation}
This relation provides a boundary condition connecting light leaving the effective FPI with light entering it, and the intracavity dynamics. It can be identified with the familiar relation $\hat{a}_\text{out}-\hat{a}_\text{in}= \sqrt{2\kappa(\hat{x})}$ taking $\kappa(\hat{x})=\kappa_\text{c}+g_\kappa\hat{x}$ and truncating the square-root to first order in $\hat{x}$.
\par
The generalized optomechanical Hamiltonian~\eqref{eq:HOM} is realized in the MSI shown in~\fref{fig:PhysicalModel} operating close to a dark-port condition, \ie, when most light is directly reflected at the first beam splitter (BS). The whole system can then be described as an effective Fabry--P\'erot interferometer of length $L$, operating in the good-cavity limit, formed between the perfect end-mirror $\mathrm{M}_1$ and an \textit{effective} end-mirror $\mathcal{M}$; $2L$ is the length of the Sagnac mode $\mathrm{M}_1$--$\mathrm{BS}$--$\mathrm{M}_2$--$\mathrm{M}_3$--$\mathrm{BS}$. The reflectivity $\rho$ and transmissivity $\tau$ of $\mathcal{M}$ depend on the complex reflectivity $R$ ($\mathfrak{r}$) and transmissivity $T$ ($\mathfrak{t}$) of $\mathrm{BS}$ (of the micromirror), as well as the displacement $\delta$ of $\mathrm{M}$ from the mid-point of $\mathrm{M}_2$--$\mathrm{M}_3$:
\begin{subequations}
\begin{align}
\rho&=-\bigl(R^2\mathfrak{r}e^{2i\delta}+T^2\mathfrak{r}e^{-2i\delta}+2RT\mathfrak{t}\bigr)e^{-i\arg{\mathfrak{t}}}\,,\text{ and}\\
\tau&=\bigl[\bigl(R{T^\ast}e^{2i\delta}-\mathrm{c.c.}\bigr)\mathfrak{r}-(\abs{R}^2-\abs{T}^2)\mathfrak{t}\bigr]e^{-i\arg{\mathfrak{t}}}\,.
\end{align}
\end{subequations}
Assuming the close-to-dark-port condition $\abs{\tau}\ll1$, a quantization along the standard routes of cavity QED~(see Appendix) yields a Hamiltonian of the form in \eref{eq:HOM} with the usual result $\kappa_\text{c}=-c/(2L)\ln\abs{\rho}\approx c\abs{\tau}^2/(4L)$, as well as dispersive and dissipative couplings
\begin{subequations}
\begin{align}
g_\omega&=\bigl(\omega_\text{c} x_0/L\bigr)\bigl[\bigl(\abs{R}^2-\abs{T}^2\bigr)+\tau\cos(\arg{\mathfrak{t}})\bigr]\,,\\
g_\kappa&=i\abs{\tau}e^{i\phi}\bigl(\omega_\text{c} x_0/L\bigr)\bigl[2RT+\rho\cos(\arg{\mathfrak{t}})\bigr]\,,
\end{align}
\end{subequations}
where $\phi\approx0$ close to resonance. These results guarantee that the values of $g_\omega$ and $g_\kappa$ can be controlled independently by choosing $\delta$ (\ie, positioning the membrane) and the reflectivity of the central beam splitter $\mathrm{BS}$ appropriately. The need for a sharp resonance demands $\abs{R}\approx\abs{T}$. We note, however, that $\abs{R}\neq\abs{T}$ is required to be able to set $g_\omega=0$ with $\abs{\tau}>0$. In the following we will specialize to this most interesting case of a \emph{purely} dissipative coupling. We will provide experimental case studies below and show that strong coupling, where $\abs{\overline{a}}\abs{g_\kappa}\gg\kappa_\text{c}$ ($\overline{a}$ being the intracavity amplitude), can be achieved for a moderate driving power of a few hundred-$\upmu$W.
\par
The next step in our investigation is to derive the Heisenberg equations of motion for $\hat{a}$, $\hat{x}$, and $\hat{p}=(\hat{b}-\hat{b}^\dagger)/\bigl(i\sqrt{2}\bigr)$. The Hamiltonian~\eqref{eq:HOM} implies a full and rich nonlinear dynamics~\cite{Marquardt2006}, including such effects as bistability and self-induced oscillations, but in this new context of dissipative optomechanics. Here we will focus on the linear dynamics by assuming a strong classical driving field $\overline{a_\text{in}}$, which allows us to write the linear equations of motion for the fluctuations around the steady state as~(see Appendix)
\begin{figure}[t]
 \includegraphics[width=\figurewidth]{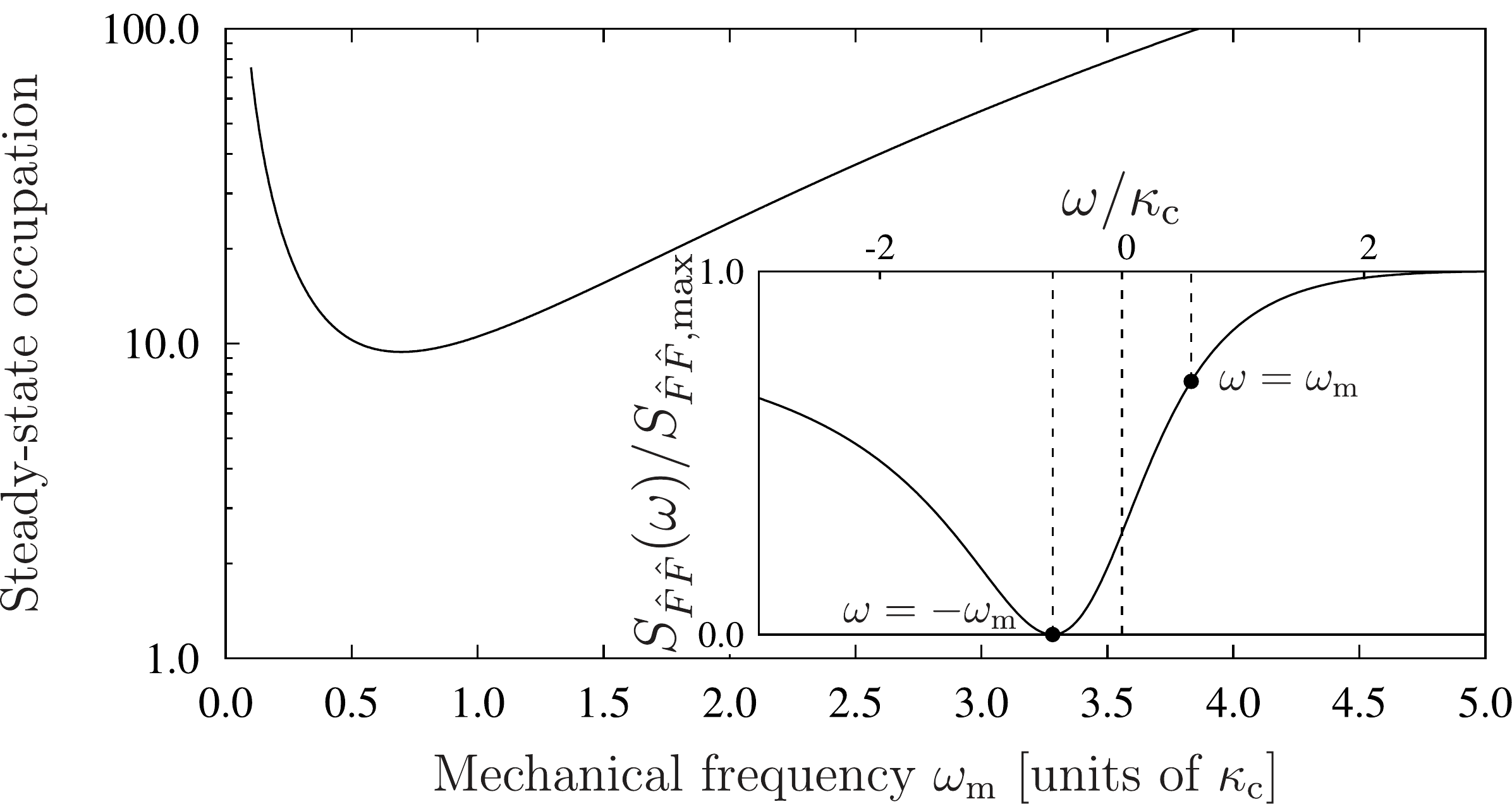}
 \caption{Calculated steady-state occupation number for system II. Throughout this plot, $\Delta=\omega_\text{m}/2$ and $P_\text{in}=10$\,nW. Inset:\ Normalized backaction force noise density, illustrating its Fano profile, when $\omega_\text{m}\approx0.6\kappa_\text{c}$ and $\Delta=\omega_\text{m}/2$; the noise density at $\omega=-\omega_m$ is zero.}
 \label{fig:OptimalomegamScan}
\end{figure}
\begin{align}
\dot{\hat{a}}&=\bigl(i\Delta-\kappa_\text{c}\bigr)\hat{a}-\sqrt{2\kappa_\text{c}}\hat{a}_\text{in}-g_\kappa\bigl(\overline{a_\text{in}}/\sqrt{2\kappa_\text{c}}+\overline{a}\bigr)\hat{x}\,,\\
\label{eq:EoMp}
\dot{\hat{p}}&=-\omega_\text{m}\hat{x}-2\kappa_\text{m}\hat{p}-\sqrt{2\kappa_\text{m}}\,\hat{\xi}\nonumber\\
&\qquad-ig_\kappa/\sqrt{2\kappa_\text{c}}\bigl[\bigl(\overline{a_\text{in}}^\ast\hat{a} + \overline{a}\,\hat{a}_\text{in}^\dagger\bigr)-\mathrm{H.c.}\bigr]\,,
\end{align}
and $\dot{\hat{x}}=\omega_\text{m}\hat{p}$, with the mechanical motion having damping rate $\kappa_\text{m}$. $\overline{a}$ is the coherent part of the cavity field; and $\hat{\xi}=\hat{\xi}^\dagger$, which models Brownian-motion--type noise acting on the mechanical oscillator, is assumed to obey $\expt{\hat{\xi}(t)\hat{\xi}(t^\prime)}=(2n_\text{th}+1)\delta(t-t^\prime)$, where $n_\text{th}$ is the thermal occupation number in the absence of driving, and $\expt{\hat{\xi}}=0$. The condition for the c-number component $\overline{a_\text{in}}$ of the input field to be `large enough' is obtained by evaluating the contributions of both components of the input field to $\expt{\hat{a}^\dagger\hat{a}}$. This linearization condition can be stated as $\abs{\overline{a_\text{in}}}^2\gg\bigl[(\abs{\Delta}+\omega_\text{m})^2+\kappa_\text{c}^2\bigr]/(2\kappa_\text{c})$~(see Appendix), and will be assumed to be satisfied in the following. In the weak-coupling limit, when $\abs{g_\kappa}^2\lvert\overline{a_\text{in}}\rvert^2\ll4\kappa_\text{c}^3$, we assume that the cavity field follows the mechanical motion adiabatically and solve the linearized dynamics to obtain the steady-state mechanical occupation number:
\begin{align}
\expt{\hat{b}^\dagger\hat{b}}=&\ \frac{n_\text{th}\kappa_\text{m}}{\bar{\kappa}_\text{m}}+\frac{\abs{g_\kappa}^2}{4\bar{\kappa}_\text{m}}\frac{\lvert\overline{a_\text{in}}\rvert^2}{\Delta^2+\kappa_\text{c}^2}\nonumber\\
&\ \times\frac{\kappa_\text{c}\bigl(2\Delta-\bar{\omega}_\text{m}\bigr)^2+\bar{\kappa}_\text{m}\bigl(\Delta^2+\kappa_\text{c}^2+\kappa_\text{c}\bar{\kappa}_\text{m}\bigr)}{\kappa_\text{c}\bigl[(\Delta-\bar{\omega}_\text{m})^2+(\kappa_\text{c}+\bar{\kappa}_\text{m})^2\bigr]}\,.
\end{align}
In this equation $\bar{\omega}_\text{m}$ and $\bar{\kappa}_\text{m}$ are the optically-shifted mechanical oscillator frequency and damping rate, respectively, whose expressions are rather involved and will not be reproduced here~(see Appendix). The preceding relation is valid even for moderately strong input powers, as can be shown by comparing this analytical result with that obtained by solving the above equations of motion exactly using the method in Ref.~\cite{Vitali2007}. In the `cryogenic optomechanics' limit, where $\kappa_\text{c}\gg\bar{\kappa}_\text{m}\gg n_\text{th}\kappa_\text{m}$ and $\bar{\omega}_\text{m}\approx\omega_\text{m}$, this result simplifies to
\begin{align}
\expt{\hat{b}^\dagger\hat{b}}\approx \frac{\abs{g_\kappa}^2}{4\bar{\kappa}_\text{m}}\frac{\lvert\overline{a_\text{in}}\rvert^2}{\Delta^2+\kappa_\text{c}^2}\frac{(2\Delta-\omega_\text{m})^2}{(\Delta-\omega_\text{m})^2+\kappa_\text{c}^2}\,.
\end{align}
This expression for $\expt{\hat{b}^\dagger\hat{b}}$ deserves some comments. We recall that in dispersive optomechanics, cooling is optimized when the upper motional sideband is strongly enhanced ($\Delta=-\omega_\text{m}$) and the lower sideband strongly suppressed ($\omega_\text{m}\gg\kappa_\text{c}$). In the present case, Fano--like interference can be observed in the backaction force noise spectrum $S_{\hat{F}\hat{F}}(\omega)$, cf.\ the inset of \fref{fig:OptimalomegamScan}, where $\hat{F}$ [the second line in \eref{eq:EoMp}] is the force operator acting on the micromirror motion. This resonance neutralizes the lower sideband when $\Delta=\omega_\text{m}/2$; the Fano lineshape also means that optimal enhancement of the upper sideband requires not $\omega_\text{m}\gg\kappa_\text{c}$---indeed, that would be disingenuous---but $\omega_\text{m}=\omega_\text{m}^\text{opt}=\frac{2}{3}\sqrt{\frac{2\sqrt{13}-5}{3}}\kappa_\text{c}\approx0.6\kappa_\text{c}$, which is much less demanding than the sideband-resolved condition. \fref{fig:OptimalomegamScan} shows how the mechanical occupation number, at the optimal detuning $\Delta=\omega_\text{m}/2$ and with an input power $P_\text{in}=10$\,nW, changes as the ratio $\omega_\text{m}/\kappa_\text{c}$ is varied, and is minimized when $\omega_\text{m}\approx\omega_\text{m}^\text{opt}$.

\begin{figure}[t]
 \centering
 \rlap{(a)}{\vspace{-1.2em}
 \includegraphics[width=\figurewidth]{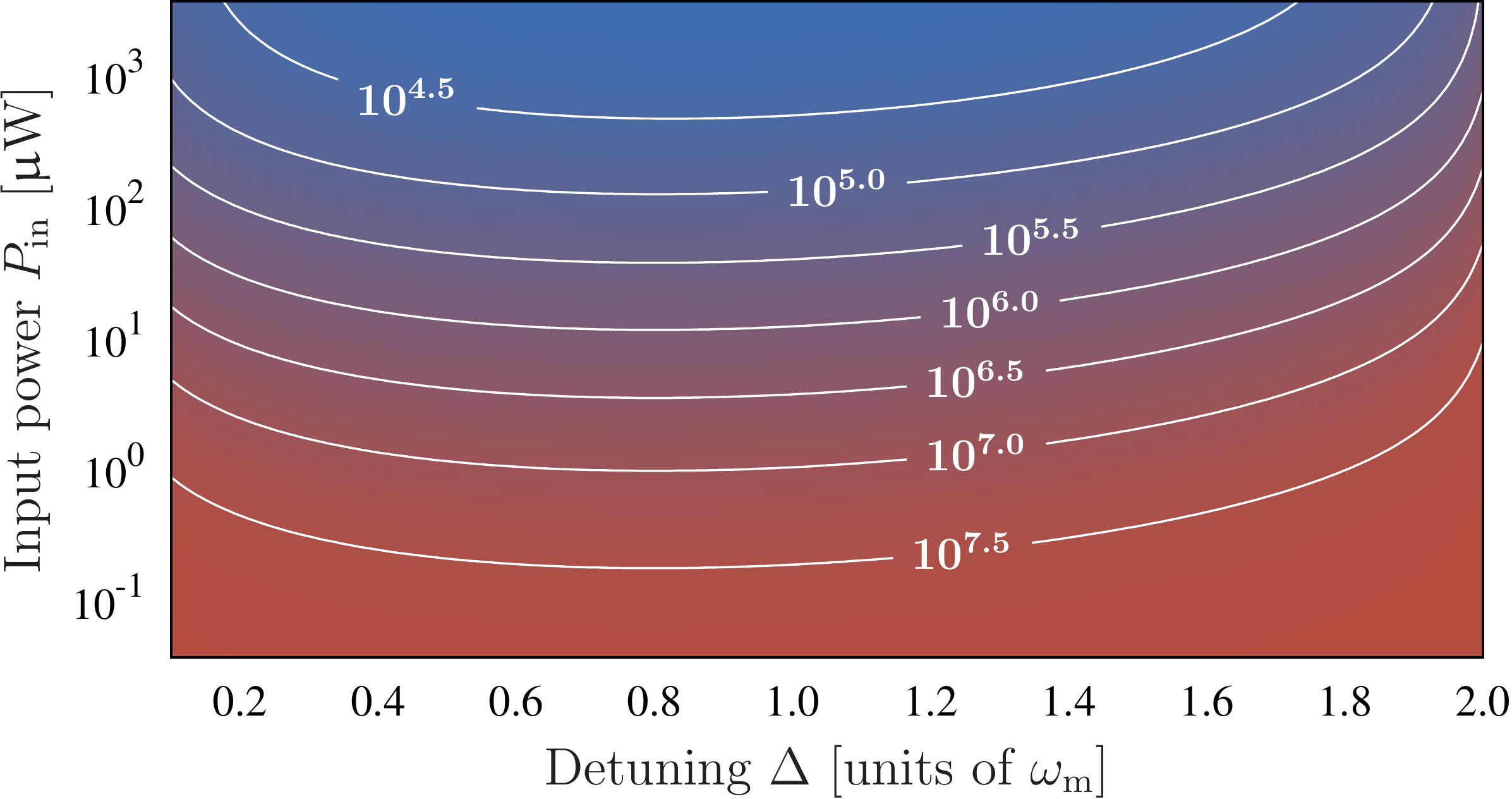}
 }\\
 \rlap{(b)}{\vspace{-1.2em}
 \includegraphics[width=\figurewidth]{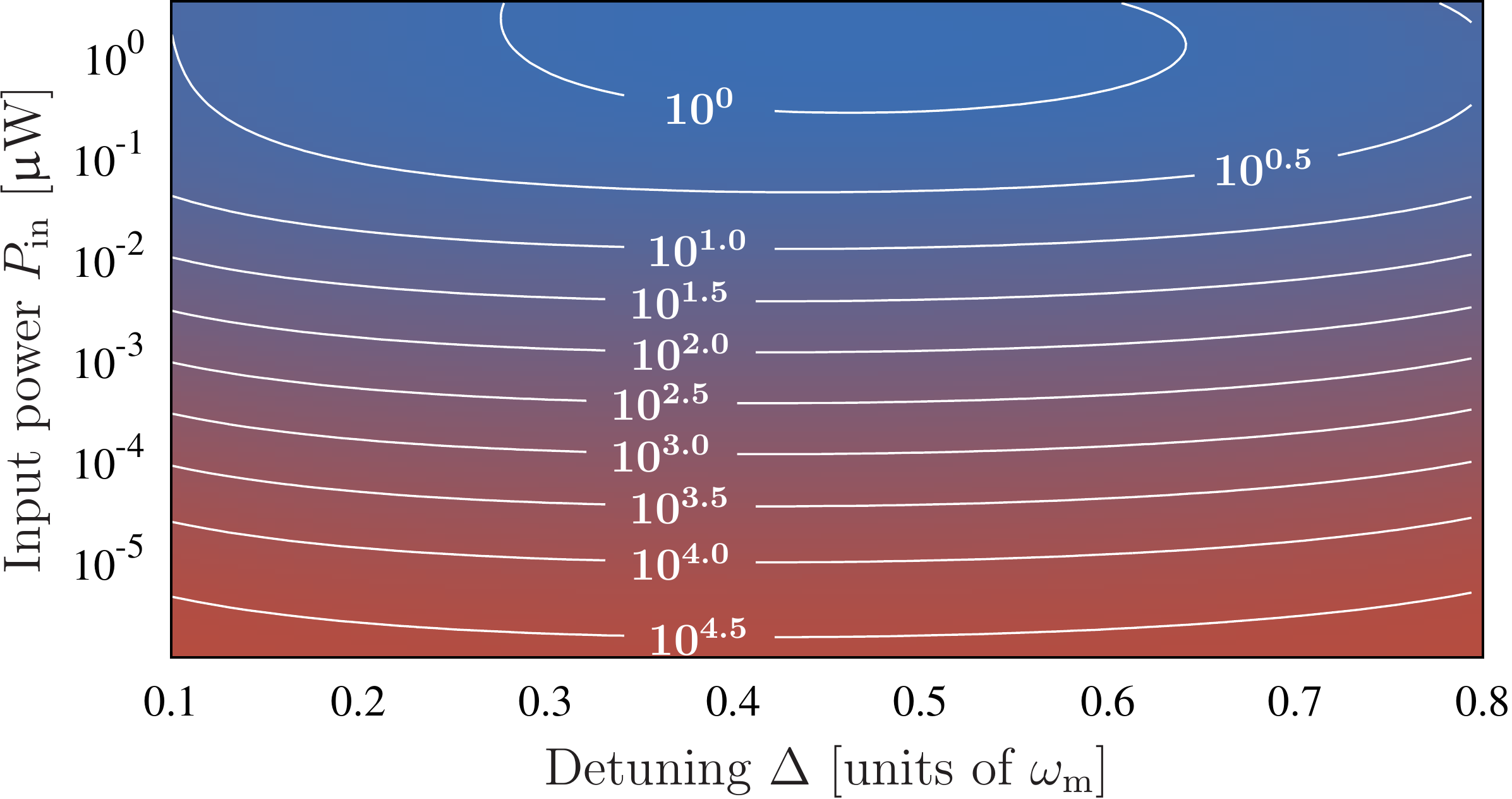}
 }
 \caption{(Color online) Full numerical solution for the occupation number $\expt{\hat{b}^\dagger\hat{b}}$ for (a)~system I, and (b)~system II.}
 \label{fig:OccupationNumberContourPlots}
\end{figure}
\section{Cooling}
Let us now turn our attention towards predicting the cooling performance of the model investigated above. We shall call `system I' the following:\ mechanical oscillator effective mass $m=100$\,ng, frequency $\omega_\text{m}=2\pi\times103$\,kHz, and quality factor $Q=\omega_\text{m}/(2\kappa_\text{m})=2\times10^6$, such that the zero-point fluctuation is $x_0=\sqrt{\hbar/(2m\omega_\text{m})}\simeq 1~\textrm{fm}$; $\abs{R}^2/\abs{T}^2=0.486/0.514$; $\abs{\mathfrak{r}}^2/\abs{\mathfrak{t}}^2=0.362/0.638$; driving wavelength $\lambda_\text{c}=1064$\,nm; and $L=7.5$\,cm; which are experimentally realizable~\cite{Friedrich2011} and yield $\kappa_\text{c}=2\pi\times196$\,kHz, $g_\kappa\approx-2\pi\times0.1$\,Hz ($g_\kappa/x_0\approx-2\pi\times79$\,kHz/nm). We limit the input power to the regime where heating from the power absorbed is not the dominant process. An input power of $10$\,mW corresponds to an effective coupling strength $G=\abs{\overline{a}}\abs{g_\kappa}\approx0.1\,\kappa_\text{c}$. By starting from an environment temperature $T_\text{env}=300$\,K we can decrease the occupation number by over three orders of magnitude, as illustrated in \fref{fig:OccupationNumberContourPlots}(a).\\
Consider next a hypothetical, but still physically realizable, situation (`system II') where $\mathrm{M}$ has a smaller mass ($m=50$\,pg~\cite{Thompson2008}), higher mechanical quality ($Q=1.1\times10^7$~\cite{Zwickl2008}), and higher reflectivity ($\abs{\mathfrak{r}}^2/\abs{\mathfrak{t}}^2=0.818/0.182$, possible by patterning the membrane~\cite{Kemiktarak2011}), where $\mathrm{BS}$ is more balanced ($\abs{R}^2/\abs{T}^2=0.496/0.504$, $\kappa_\text{c}=2\pi\times59$\,kHz), and thus $g_\kappa\approx-2\pi\times2.6$\,Hz ($g_\kappa/x_0\approx-2\pi\times65$\,kHz/nm). For this system, we can achieve the strong-coupling condition $G\gtrsim\kappa_\text{c}$. If we also assume cryogenic operation at $T_\text{env}=0.3$\,K, we can see from \fref{fig:OccupationNumberContourPlots}(b) that ground-state cooling is possible, despite the poor reflectivity of the mechanical oscillator, and that $\omega_\text{m}\sim\kappa_\text{c}$.

\emph{Position transduction.}---Optomechanical systems are one promising approach towards extremely sensitive position transduction~\cite{Kippenberg2008}. The coupling of a mechanical oscillator to multiple cavity modes was recently explored in Ref.~\cite{Dobrindt2010}; a common feature of `multiple light mode--single mechanical mode' systems is a Fano-like profile in $S_{\hat{F}\hat{F}}(\omega)$ (\fref{fig:OptimalomegamScan} inset). The corresponding anti-resonance allows one to reach the SQL for measurement imprecision at a significantly lower input power than a `single light mode--single mechanical mode' optomechanical system. Indeed, let us quantify the achievable resolution of a position measurement by the ratio $\mathcal{N}/\mathcal{S}$, where the `noise' $\mathcal{N}^2$ is the contribution to the symmetrized homodyne output spectrum evaluated at $\omega=\omega_\text{m}$, $\overline{S}_\text{out}(\omega_\text{m})$, due to $\hat{a}_\text{in}$, and the `signal' $\mathcal{S}^2$ is that due to $\hat{\xi}$, normalised to the free mechanical motion noise spectrum $\overline{S}_\text{free}(\omega_\text{m})$:
\begin{equation}
\overline{S}_\text{out}(\omega_\text{m})=\mathcal{N}^2+\mathcal{S}^2\,\overline{S}_\text{free}(\omega_\text{m})\,.
\end{equation}
At $\Delta=0$ and under identical conditions, it can be shown that $\mathcal{N}/\mathcal{S}$ reaches the same lower bound (the SQL) in both dissipative (at a power $P_\kappa$) and dispersive ($P_\omega$) cases, but with $P_\kappa=(2\kappa/\omega_\text{m})^2P_\omega$. In the sideband-resolved regime, therefore, one can obtain \emph{significantly better position resolution} at low powers in comparison with the dispersive case.
\par
\section{Comments}
We set $g_\omega=0$ early on in this paper, which can be realized by placing $\mathrm{M}$ at a point where the field intensity surrounding it is close to minimal, greatly reducing the power absorbed $P_\text{abs}$ by the membrane. For the parameters in \fref{fig:OccupationNumberContourPlots}(a), using the `thermal link' from Ref.~\cite{Hammerer2009}, the membrane temperature rises by ca.~$60$\,K at $P_\text{in}=1$\,mW. The resulting temperature rise has a significant effect on the base occupation number of the micromirror, but still allows strong cooling of the micromirror motion. It is a feature of our topology that the ideal position of $\mathrm{M}$ corresponds to both where $P_\text{abs}$ is greatly reduced, and where the competing dispersive optomechanics is switched off.
\par
Significantly, we note that this situation does not persist in the case of the `membrane-in-the-middle' geometry. In a single-transverse-mode model for this latter situation, the dispersive and dissipative optomechanics cannot be independently turned off, and are both zero at the nodes of the cavity field; this leads to a stronger restriction arising from the power absorbed. Moreover, dispersive (dissipative) optomechanical cooling requires $\Delta<0$ ($\Delta>0$); most treatments of this geometry do not include the dissipative optomechanics component of the dynamics~\cite{Jayich2008}, and (at least within a single-mode model) may therefore over-estimate the cooling efficiency in the non-sideband-resolved regime.
\par
\section{Conclusions}
In this paper we have presented an experimentally feasible realization of an optomechanical system that can be fully tuned between strong dissipative and dispersive dynamics. The cooling mechanism we presented works best in the non-sideband-resolved regime, unlike the usual dispersive case. For an existing set of experimental parameters, we predict a strong cooling effect arising from a Fano--type resonance in the backaction force acting on the mechanical motion. In the case of more optimistic parameters, we predict ground-state cooling of the mechanical motion. In the opposite, sideband-resolved, regime, our system promises significantly improved position measurement resolution. The usage of the proposed implementation of a dissipative optomechanical coupling for nonlinear (quantum) dynamics~\cite{Nunnenkamp2011,Rabl2011}, and for pulsed dynamics~\cite{Vanner2010} remain to be explored.
\\
Moreover, we believe that the topology of a MSI with a highly reflective mirror in its dark port---the signal recycling configuration used in the context of gravitational wave detectors~\cite{Meers1988,Willke2002}---and the resulting dissipative intracavity dynamics provides attractive possibilities for other fields in cavity QED, especially in combination with single atoms or ensembles of cold atoms.

\acknowledgments
We thank H.\ M\"uller-Ebhardt for useful discussions. This work was funded by the Centre for Quantum Engineering and Space-Time Research (QUEST) at the Leibniz University Hannover.

\appendix
\begin{figure}[t]
 \centering
  \includegraphics[scale=0.9]{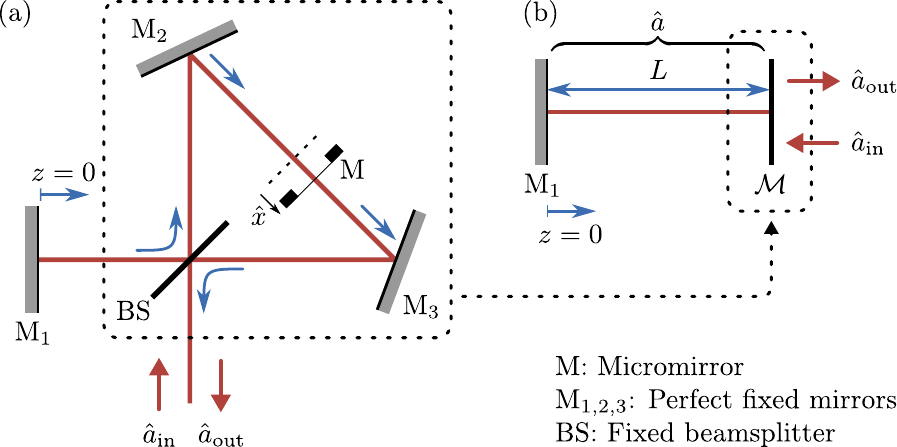}
\caption{(a)~Schematic layout of the Michelson--Sagnac interferometer. The $z$-axis is defined as having its origin at $\mathrm{M}_1$. $\hat{x}$ is the operator representing the displacement of $\mathrm{M}$ from its equilibrium position. (b)~The optics in the dotted box in (a) form an effective mirror, $\mathcal{M}$.}
 \label{fig:Interferometer}
\end{figure}
\begin{figure}
 \centering
 \includegraphics[scale=0.9]{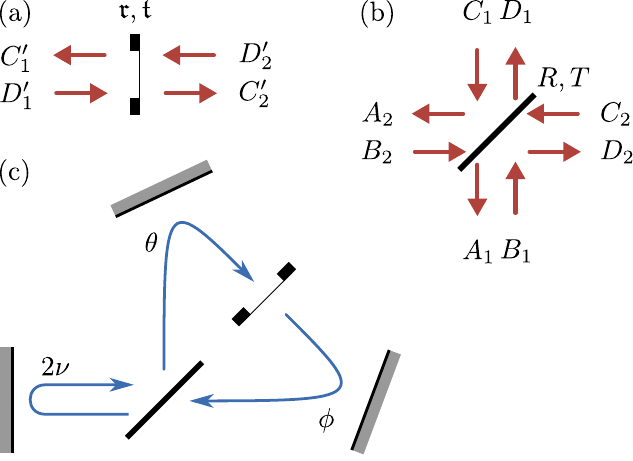}
\caption{(a)~The four field modes interacting with the micromirror. (b)~The eight field modes interacting with the beamsplitter. $C_{1,2}$ and $D_{1,2}$ are identical, up to a phase shift, to $C_{1,2}^\prime$ and $D_{1,2}^\prime$, respectively. (c)~Phase shifts incurred in each of the three sections of the interferometer, at the cavity frequency $\omega_\text{c}$.}
 \label{fig:Fields}
\end{figure}
\section{Field inside the interferometer}\label{sec:Amplitudes}
Consider the Michelson--Sagnac interferometer topology shown in \fref{fig:Interferometer}, where all the fixed mirrors are perfect.
The micromirror, assumed lossless, has amplitude reflectivity $\mathfrak{r}$ and transmissivity $\mathfrak{t}$.
The corresponding parameters for the lossless beamsplitter are $R$ and $T$, respectively.
The electric field for the half-space defined by $z\geq0$ can be written as
\begin{equation}
E^+(z)=\int_0^\infty\rmd\omega\mathcal{N}(\omega)u_\omega(z)\hat{a}_\omega\,,
\end{equation}
where the mode function
\begin{equation}
u_\omega(z)=\begin{cases}
A_2e^{-ikz}+B_2e^{ikz}&0\leq z\leq z_1\\
C_1e^{-ikz}+D_1e^{ikz}&z_1\leq z\leq z_2\\
D_2e^{-ikz}+C_2e^{ikz}&z_2\leq z\leq z_3\\
B_1e^{-ikz}+A_1e^{ikz}&z_3\leq z
\end{cases}\,,
\end{equation}
with reference to \fref{fig:Fields}.
Starting from $z=0$, $\mathrm{BS}$ is at $z=z_1$, $\mathrm{M}$ at $z=z_2$, then $\mathrm{BS}$ is encountered again at $z=z_3$.
For a beam of cross-sectional area $\mathcal{A}$, the frequency-dependent normalization factor is $\mathcal{N}(\omega)=\sqrt{\frac{\hbar\omega}{4\pi\mathcal{A}\epsilon_0c}}$.
With $\hat{a}_\omega$ being the annihilation operator for the mode at frequency $\omega$, we can thus write the field Hamiltonian ($\hbar=1$)
\begin{equation}
\hat{H}=\int\rmd\omega\,\omega\,\hat{a}_\omega^\dagger\hat{a}_\omega\,.
\end{equation}
In this half-space, we can solve for all the field mode amplitudes, $A_{1,2}$, $B_2$, $C_{1,2}$, $C^\prime_{1,2}$, $D_{1,2}$, and $D^\prime_{1,2}$, in terms of $B_1$ by, \eg, using the transfer matrix formalism~\cite{Deutsch1995,Xuereb2009b}.
Upon writing this solution, it becomes apparent that the optics inside the dotted box in \fref{fig:Interferometer} can be treated as a single effective mirror $\mathcal{M}$, having reflectivity and transmissivity
\begin{align}
\rho&=-\bigl(R^2\mathfrak{r}e^{i\delta}+T^2\mathfrak{r}e^{-i\delta}+2RT\mathfrak{t}\bigr)e^{i(\nu-\chi)}\,,\text{ and}\\
\tau&=\bigl[R{T^\ast}\mathfrak{r}e^{i\delta}-R^\ast{}T\mathfrak{r}e^{-i\delta}-(\abs{R}^2-\abs{T}^2)\mathfrak{t}\bigr]e^{i(\nu-\chi)}\,,
\end{align}
where $\nu$ is the phase-delay $\mathrm{M}_1$--$\mathrm{BS}$, $\delta=\theta-\phi$ is the arm-length difference for the loop in the interferometer, cf.\ \fref{fig:Fields}(c), and $\chi=\arg(\mathfrak{t})$.
In this notation, we can now write, \eg,
\begin{equation}
\label{eq:B2Lorentzian}
B_2=\frac{\tau\exp[i(\sigma+\chi+\nu)]}{1-\rho\exp[i(\sigma+\chi+\nu)]}B_1;
\end{equation}
for convenience we have defined $\sigma=\theta+\phi$. The coefficient linking the ``intracavity'' field amplitude $B_2$ to the one of the incoming field $B_1$ is completely equivalent to the familiar spectral response function of a Fabry-Perot cavity, see \eg\ Ref.~\cite{Vogel2006}.

\section{Interaction Hamiltonian}\label{sec:RPHamiltonianThin}
In this section, we will derive an effective Hamiltonian $\hat{H}_\text{int}$ that describes the interaction between the motion of the micromirror and the electric field in the regime where the displacement $\hat{x}$ of the membrane from its average position (fixed by $\delta$) is small on the scale of a wave length, \textit{i.e.} in the Lamb--Dicke regime. The Hamiltonian for the radiation-pressure interaction in linear order of $\hat{x}$ is~\cite{Hammerer2010}
\begin{multline}
\label{eq:HintThick}
\hat{H}_\text{int}=\mathcal{A}\epsilon_0(\eta^2-1)\bigl[E^-\bigl(z_2+\tfrac{d}{2}\bigr)E^+\bigl(z_2+\tfrac{d}{2}\bigr)\\
-E^-\bigl(z_2-\tfrac{d}{2}\bigr)E^+\bigl(z_2-\tfrac{d}{2}\bigr)\bigr]\hat{x}\,,
\end{multline}
with membrane thickness $d$ and refractive index $\eta$.
We now make the simplification that the membrane is infinitesimally thin, which corresponds to taking $d\to0$ and $\eta\to\infty$ such that $\eta^2d$ is constant.
This allows us to express $\hat{H}_\text{int}$ in terms of left (right) propagating fields, $E^\pm_{L(R)}$, through $u_{\text{L}(\text{R}),\omega}(z)$:
\begin{subequations}
\begin{align}
u_{\text{L},\omega}(z_2^-)&=\bigl(\mathfrak{r}Re^{i\theta}+\mathfrak{t}Te^{i\phi}\bigr)B_1\nonumber\\
&\qquad-\bigl(\mathfrak{r}R^\ast e^{i\theta}-\mathfrak{t}T^\ast e^{i\phi}\bigr)e^{2i\nu}B_2\\
u_{\text{R},\omega}(z_2^-)&=Re^{i\theta}B_1+R^\ast e^{i(2\nu+\theta)}B_2\\
u_{\text{L},\omega}(z_2^+)&=Te^{i\phi}B_1-T^\ast e^{i(2\nu+\phi)}B_2\\
u_{\text{R},\omega}(z_2^+)&=\bigl(\mathfrak{t}Re^{i\theta}+\mathfrak{r}Te^{i\phi}\bigr)B_1\nonumber\\
&\qquad+\bigl(\mathfrak{t}R^\ast e^{i\theta}-\mathfrak{r}T^\ast e^{i\phi}\bigr)e^{2i\nu}B_2\,,
\end{align}
\end{subequations}
which are expressed in terms of $B_1$ and $B_2$ in anticipation of taking the good-cavity limit, and where $z_2^\pm=\lim_{d\to0}\bigl(z_2\pm\tfrac{d}{2}\bigr)$.
With this notation, we have (cf.\ Ref.~\cite{Xuereb2009b})
\begin{multline}\label{eq:Hint}
\hat{H}_\text{int}=2\mathcal{A}\epsilon_0\bigl[E^-_\text{L}\bigl(z_2^-\bigr)E^+_\text{L}\bigl(z_2^-\bigr)+E^-_\text{R}\bigl(z_2^-\bigr)E^+_\text{R}\bigl(z_2^-\bigr)\\
-E^-_\text{L}\bigl(z_2^+\bigr)E^+_\text{L}\bigl(z_2^+\bigr)-E^-_\text{R}\bigl(z_2^+\bigr)E^+_\text{R}\bigl(z_2^+\bigr)\bigr]\hat{x}\,.
\end{multline}

\section{Good-cavity limit}\label{sec:GoodCavity}
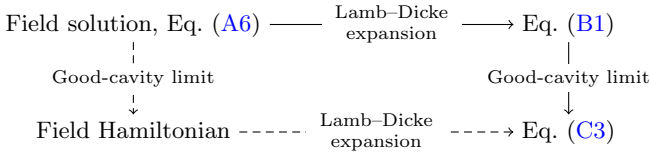
\begin{figure}[t]
 \centering
\begin{tikzpicture}[description/.style={fill=white,inner sep=2pt}]
\matrix (m) [matrix of math nodes, row sep=3em, column sep=10em, text height=1.5ex, text depth=0.25ex]{
\small\text{Field solution, \eref{eq:B2Lorentzian}} & \text{\eref{eq:HintThick}} \\
\small\text{Field Hamiltonian} & \text{\eref{eq:HRPCompact}} \\
};
\path[->,font=\scriptsize]
(m-1-1) edge node[description] {$\begin{array}{c}\text{Lamb--Dicke}\\\text{expansion}\end{array}$} (m-1-2)
(m-1-2) edge node[description] {Good-cavity limit} (m-2-2);
\path[densely dashed,->,font=\scriptsize]
(m-1-1) edge node[description] {Good-cavity limit} (m-2-1)
(m-2-1) edge node[description] {$\begin{array}{c}\text{Lamb--Dicke}\\\text{expansion}\end{array}$} (m-2-2);
\end{tikzpicture}
\caption{Schematic of the alternative paths to obtaining the interaction Hamiltonian. Here, we follow the clockwise path shown by the solid arrows, but an equivalent formulation follows the dashed arrows.}
 \label{fig:AltPaths}
\end{figure}
We now go to the good-cavity limit in our system.
This condition, which is often termed the `dark port' (or `dark fringe') condition~\cite{Yamamoto2010}, corresponds to having $\abs{\rho}\approx 1$.
Under such conditions, the system acts as an effective Fabry--P\'erot interferometer with one perfect end-mirror, $\mathrm{M}_1$, and the effective end-mirror $\mathcal{M}$.
In other words, $\abs{B_2}^2$ exhibits a Lorentzian behavior, with the peaks occurring at the poles of the denominator of \eref{eq:B2Lorentzian}, at
\begin{equation}
\omega_\text{c} = \frac{c}{2L}\bigl[\arg(\rho)-(\sigma+\chi+\nu)+2\pi{}n\bigr]\,,
\end{equation}
with $n$ being an integer, and having a half-width at half-maximum (HWHM) $\kappa_\text{c} = -c/(2L)\ln\abs{\rho}$, $L=(\sigma+\nu)c/(2\omega_\text{c})$ being the length of the interferometer.
\par
In the following we will concentrate on one particular cavity resonances, and consider $n$ to be fixed. For this single mode cavity field we introduce the bosonic field annihilation operators $\hat{a}$ and reassign $\hat{a}_\omega$ to be the annihilation operators for the bath that the cavity field couples to; these operators satisfy the usual commutation relations
\begin{equation}
\bigl[\hat{a}_\omega,\hat{a}^\dagger_{\omega^\prime}\bigr]=\delta(\omega-\omega^\prime)\,,\text{ and }\bigl[\hat{a},\hat{a}^\dagger\bigr]=1\,.
\end{equation}
We can then write the interaction Hamiltonian \eqref{eq:Hint} as
\begin{multline}
\label{eq:HRPCompact}
\hat{H}_\text{int}=g_0\bigl(\alpha/\sqrt{2}\bigr)\hat{a}^\dagger\hat{a}\,(\hat{b}+\hat{b}^\dagger)\\
+g_0\sqrt{\frac{L}{2\pi{}c}}\biggl(\beta\int\rmd\omega\,\hat{a}_\omega^\dagger\,\hat{a} + \mathrm{H.c.}\biggr)(\hat{b}+\hat{b}^\dagger)\,,
\end{multline}
where for clarity we have defined
\begin{align}
\alpha&=2\bigl[\bigl(\abs{R}^2-\abs{T}^2\bigr)+\tau{}e^{-i\nu}\cos(\chi)\bigr]\,,\text{ and}\\
\beta&=-2\bigl[2RT+\rho{}e^{-i\nu}\cos(\chi)\bigr]e^{i(\sigma+\chi+\nu)}\,,
\end{align}
We may relate $\alpha$ and $\beta$ to $g_{\omega,\kappa}$ from the main text by:
\begin{equation}
g_\omega=g_0\bigl(\alpha/\sqrt{2}\bigr)\,,\text{ and }g_\kappa=-i g_0\beta\sqrt{\frac{4\kappa_c L}{c}}\approx-ig_0\abs{\tau}\beta\,.
\end{equation}
which are assumed to be slowly varying over the frequency range of interest.
The coupling constant $g_0$, which is dimensionally a frequency, is defined as $g_0=\omega_\text{c}x_0/2L$, and $\hat{b}$ represents the annihilation operator for the motion of the membrane.
In terms of $\hat{b}$ and the zero-point fluctuations $x_0$, we can write the displacement of the micromirror from its equilibrium position as $\hat{x}=x_0(\hat{b}+\hat{b}^\dagger)/\sqrt{2}$.
(Note that in the main text we used a dimensionless $\hat{x}$.)
\par
In the above formulation, we first computed the Hamiltonian to first order in $\hat{x}$ and \emph{then} took the good-cavity limit.
An equivalent formulation would be to first take the good-cavity limit, with the linewidth of the cavity depending parametrically on the position of the membrane, and then expanding this linewidth to first order in $\hat{x}$ (cf.\ \fref{fig:AltPaths}). We also remark that the analysis can be performed without taking the good-cavity limit, resulting in equivalent results but less transparent expressions.

\section{Equations of motion}\label{sec:EoM}
We can express the full Hamiltonian $\hat{H}$ as a sum of the free cavity $\hat{H}_\text{c}$, free micromirror $\hat{H}_\text{m}$, free field $\hat{H}_\text{f}$, field--cavity interaction $\hat{H}_\text{f--c}$, and cavity--micromirror interaction $\hat{H}_\text{int}$ Hamiltonians:
\begin{subequations}
\label{eq:Hamiltonians}
\begin{align}
\label{eq:Hsum}
\hat{H}&=\hat{H}_\text{c}+\hat{H}_\text{m}+\hat{H}_\text{b}+\hat{H}_\text{b--c}+\hat{H}_\text{int}\,,\\
\label{eq:Hc}
\hat{H}_\text{c}&=\omega_\text{c}\hat{a}^\dagger\hat{a}\,,\\
\label{eq:Hm}
\hat{H}_\text{m}&=\omega_\text{m}\hat{b}^\dagger\hat{b}\,,\\
\label{eq:Hb}
\hat{H}_\text{f}&=\int\rmd\omega\,\omega\,\hat{a}_\omega^\dagger\hat{a}_\omega\,,\\
\label{eq:Hbc}
\hat{H}_\text{f--c}&=i\sqrt{\frac{\kappa_\text{c}}{\pi}}\int\rmd\omega\,\bigl(\hat{a}_\omega^\dagger\hat{a}-\hat{a}^\dagger\hat{a}_\omega\bigr)\,.
\end{align}
\end{subequations}
and $\hat{H}_\text{int}$ is given in \eref{eq:HRPCompact}.

\subsection{Cavity bath: Input--output relation}
We will first calculate the equation of motion for $\hat{a}_\omega$:
\begin{equation}
\label{eq:EoMb}
\dot{\hat{a}}_\omega=-i\omega\hat{a}_\omega+\Biggl[\sqrt{\frac{\kappa_\text{c}}{\pi}}-ig_0\beta\sqrt{\frac{L}{4\pi{}c}}(\hat{b}+\hat{b}^\dagger)\Biggr]\hat{a}\,.
\end{equation}
This equation can be formally solved (see Ref.~\cite[\S 5.3]{Gardiner2004}) and, in the weak-coupling approximation, integrated over frequency to give
\begin{equation}
\label{eq:abbinrelation}
\frac{1}{\sqrt{2\pi}}\int\rmd\omega\,\hat{a}_\omega=\hat{a}_\text{in}+\Biggl[\sqrt{\frac{\kappa_\text{c}}{2}}-ig_0\beta\sqrt{\frac{L}{8c}}(\hat{b}+\hat{b}^\dagger)\Biggr]\hat{a}\,,
\end{equation}
where we have defined the input field
\begin{equation}
\label{eq:binDefinition}
\hat{a}_\text{in}\define\frac{1}{\sqrt{2\pi}}\int\rmd\omega\,e^{-i\omega(t-t_0)}\hat{a}_{\omega,0}\,,
\end{equation}
$t_0$ being some remote initial time at which $\hat{a}_\omega=\hat{a}_{\omega,0}$.
We can similarly define an output field at some distant future time $t_1$:
\begin{equation}
\label{eq:boutDefinition}
\hat{a}_\text{out}\define\frac{1}{\sqrt{2\pi}}\int\rmd\omega\,e^{-i\omega(t-t_1)}\hat{a}_{\omega,1}\,,
\end{equation}
and infer
\begin{equation}
\label{eq:abboutrelation}
\frac{1}{\sqrt{2\pi}}\int\rmd\omega\,\hat{a}_\omega=\hat{a}_\text{out}-\Biggl[\sqrt{\frac{\kappa_\text{c}}{2}}-ig_0\beta\sqrt{\frac{L}{8c}}(\hat{b}+\hat{b}^\dagger)\Biggr]\hat{a}\,.
\end{equation}
Finally, \erefs{eq:abbinrelation} and~(\ref{eq:abboutrelation}) give us the input--output relation
\begin{equation}
\label{eq:InOutReln}
\hat{a}_\text{out}-\hat{a}_\text{in}=\Biggl[\sqrt{2\kappa_\text{c}}-ig_0\beta\sqrt{\frac{L}{2c}}(\hat{b}+\hat{b}^\dagger)\Biggr]\hat{a}\,.
\end{equation}
The input field operators model a heat bath at zero temperature, and therefore satisfy the relations
\begin{subequations}
\begin{align}
\expt{\hat{a}_\text{in}(t)\hat{a}_\text{in}^\dagger(t^\prime)}&=\delta(t-t^\prime)\,,\text{ and}\\
\expt{\hat{a}_\text{in}^\dagger(t)\hat{a}_\text{in}(t^\prime)}&=0\,.
\end{align}
\end{subequations}

\subsection{Cavity field}
The equation of motion for the cavity field, correct to linear order in terms of the form $(\hat{b}+\hat{b}^\dagger)$, reads
\begin{multline}
\label{eq:EoMaUncompact}
\dot{\hat{a}}=-i\bigl[\omega_\text{c}+g_0\alpha(\hat{b}+\hat{b}^\dagger)\bigr]\hat{a}\\
-\biggl[\kappa_\text{c}+g_0\im{\beta}\sqrt{\frac{\kappa_\text{c}L}{c}}(\hat{b}+\hat{b}^\dagger)\biggr]\hat{a}\\
-\biggl[\sqrt{2\kappa_\text{c}}+ig_0\beta^\ast\sqrt{\frac{L}{2c}}(\hat{b}+\hat{b}^\dagger)\biggr]\hat{a}_\text{in}\,.
\end{multline}
Let us now define the effective cavity resonance frequency and linewidth,
\begin{subequations}
\begin{align}
\bar{\omega}_\text{c}&\define\omega_\text{c}+g_0\alpha(\hat{b}+\hat{b}^\dagger)\,,\\
\bar{\kappa}_\text{c}&\define\kappa_\text{c}+g_0\im{\beta}\sqrt{\frac{\kappa_\text{c}L}{c}}(\hat{b}+\hat{b}^\dagger)\,,
\end{align}
\end{subequations}
respectively.
Algebraic operations on $\bar{\kappa}_\text{c}$ are to be understood in terms of series expansions in powers of $\epsilon=\sqrt{\kappa_\text{c}L/c}$, correct only to first order in this parameter, \eg,
\begin{equation}
\sqrt{\bar{\kappa}_\text{c}}=\sqrt{\kappa_\text{c}}+\tfrac{1}{2}g_0\im{\beta}\sqrt{\frac{L}{c}}(\hat{b}+\hat{b}^\dagger)\,.
\end{equation}
We can simplify our expressions considerably if we introduce an effective coupling parameter $\Gamma$, which is dimensionally a frequency, through the relation
\begin{equation}
\sqrt{2\Gamma}\define{}g_0\re{\beta}\sqrt{\frac{L}{2c}}(\hat{b}+\hat{b}^\dagger)\,.
\end{equation}
where algebraic expressions involving $\Gamma$ are to be interpreted similarly to those involving $\bar{\kappa}_\text{c}$; in particular, $2\Gamma=\bigl(\sqrt{2\Gamma}\bigr)^2=0$ to first order in $\epsilon$.
In terms of these quantities, we can rewrite the input--output relation as
\begin{equation}
\hat{a}_\text{out}-\hat{a}_\text{in}=\bigl(\sqrt{2\bar{\kappa}_\text{c}}-i\sqrt{2\Gamma}\bigr)\hat{a}\,.
\end{equation}
Moreover, the equation of motion for $\hat{a}$ takes the simplified form
\begin{equation}
\label{eq:EoMa}
\dot{\hat{a}}=-(i\bar{\omega}_\text{c}+\bar{\kappa}_\text{c})\hat{a}-\bigl(\sqrt{2\bar{\kappa}_\text{c}}+i\sqrt{2\Gamma}\bigr)\hat{a}_\text{in}\,.
\end{equation}

\subsection{Micromirror motion}
The motion of the micromirror is modeled as a single-frequency harmonic oscillator. The equation of motion for $\hat{b}$ reads
\begin{multline}
\label{eq:EoMc}
\dot{\hat{b}}=-i\omega_\text{m}\hat{b}-ig_0\alpha\,\hat{a}^\dagger\hat{a}-ig_0\sqrt{\frac{L}{2c}}\bigl(\beta\,\hat{a}_\text{in}^\dagger\hat{a} + \beta^\ast\hat{a}^\dagger\hat{a}_\text{in}\bigr)\\
-ig_0\sqrt{\frac{\kappa_\text{c}L}{4c}}\bigl(\beta\,\hat{a} + \beta^\ast\hat{a}^\dagger\bigr)\,.
\end{multline}
We notice immediately that the last three terms in \eref{eq:EoMc} do not appear in the equation of motion for $\hat{x}$; they describe a Hermitian force operator
\begin{multline}
\hat{F}=-\sqrt{2}m\omega_\text{m}x_0g_0\alpha\,\hat{a}^\dagger\hat{a}\\
-m\omega_\text{m}x_0g_0\sqrt{\frac{L}{c}}\bigl(\beta\,\hat{a}_\text{in}^\dagger\hat{a} + \beta^\ast\hat{a}^\dagger\hat{a}_\text{in}\bigr)\\
-m\omega_\text{m}x_0g_0\sqrt{\frac{\kappa_\text{c}L}{2c}}\bigl(\beta\,\hat{a} + \beta^\ast\hat{a}^\dagger\bigr)\,.
\end{multline}
Defining the momentum quadrature of the micromirror motion as $\hat{p}=m\omega_\text{m}x_0(\hat{b}-\hat{b}^\dagger)/\bigl(\sqrt{2}i\bigr)$, and introducing the motion--heat-bath coupling through the Brownian-motion Langevin operator $\hat{\xi}$ we obtain $\dot{\hat{x}}=-\hat{p}/m$, and
\begin{equation}
\label{eq:Micromirrorpdot}
\dot{\hat{p}}=-m\omega_\text{m}^2\hat{x}-2\kappa_\text{m}\hat{p}+2\sqrt{\kappa_\text{m}}m\omega_\text{m}x_0\hat{\xi}+\hat{F}\,,
\end{equation}
where $\expt{\hat{\xi}}=0$ and $\expt{\hat{\xi}(t)\hat{\xi}(t^\prime)}=(2n_\text{th}+1)\delta(t-t^\prime)$, with
\begin{equation}
n_\text{th}=\frac{1}{e^{\hbar\omega_\text{m}/(k_\text{B}T_\text{env})}-1}\,
\end{equation}
being the thermal occupation number, in the absence of driving, at an environment temperature $T_\text{env}$.

\section{Linearized dynamics}
The equations of motion derived in the previous section must be linearized with respect to both $\hat{a}$ and $\hat{a}_\text{in}$.
In the case of the former, the criterion for linearization can be stated simply: $\abs{\overline{a}}^2\gg\expt{\hat{a}^\dagger\hat{a}}=1$; in the case of $\hat{a}_\text{in}$ the situation is somewhat more involved.
To see what the relevant criterion is let us examine the relevant contributions of $\hat{a}_\text{in}\hat{x}$ and $\overline{a_\text{in}}\hat{x}$ to $\expt{\hat{a}^\dagger\hat{a}}$.
Consider first, in the weak-coupling approximation,
\begin{equation}
\dot{\hat{a}}(t)=(i\Delta-\kappa_\text{c})\hat{a}(t)+\gamma\,\hat{a}_\text{in}(t)\Bigl(\hat{b}_0e^{-i\omega_\text{m}t}+\hat{b}_0^\dagger{}e^{i\omega_\text{m}t}\Bigr)\,,
\end{equation}
where we extract the harmonic time dependence, at frequency $\omega_\text{m}$, from $\hat{b}$, and where $\gamma$ is some constant.
Let us again assume that $\hat{b}_0$ varies slowly.
We can Fourier-transform this equation, compute $\expt{\hat{a}^\dagger[\omega]\hat{a}[\omega^\prime]}$, and reverse the Fourier transforms to obtain
\begin{equation}
\label{eq:ainhatLinearisationCriterion}
\expt{\hat{a}^\dagger(t)\hat{a}(t)}=\frac{\abs{\gamma}^2}{2\kappa_\text{c}}\Bigl(\expt{\hat{b}_0^\dagger\hat{b}_0}+\expt{\hat{b}_0\hat{b}_0^\dagger}\Bigr)\,.
\end{equation}
We will now look at the c-number contribution,
\begin{equation}
\dot{\hat{a}}=(i\Delta-\kappa_\text{c})\hat{a}+\gamma\,\overline{a_\text{in}}\Bigl(\hat{b}_0e^{-i\omega_\text{m}t}+\hat{b}_0^\dagger{}e^{i\omega_\text{m}t}\Bigr)\,.
\end{equation}
This equation can be solved, for the assumed case of weak coupling and slowly varying $\overline{a_\text{in}}$ and $\hat{b}_0$. Upon computing $\expt{\hat{a}^\dagger(t)\hat{a}(t)}$, we can state the linearization criterion as
\begin{equation}
\label{eq:ainLinearisationCriterion}
\abs{\overline{a_\text{in}}}^2\gg\frac{(\abs{\Delta}+\omega_\text{m})^2+\kappa_\text{c}^2}{2\kappa_\text{c}}\,.
\end{equation}
\par
Let us now calculate $\expt{\hat{b}^\dagger\hat{b}}$ in the steady state.
We start with the two coupled linearized equations of motion for $\alpha=0$ and imaginary $\beta$:
\begin{subequations}
\begin{multline}
\label{eq:EoMaSSPrecursor}
\dot{\hat{a}}=\bigl(i\Delta-\kappa_\text{c}\bigr)\hat{a}-\sqrt{2\kappa_\text{c}}\hat{a}_\text{in}\\
+ig_0\beta\sqrt{\frac{L}{2c}}\bigl(\overline{a_\text{in}}+\sqrt{2\kappa_\text{c}}\overline{a}\bigr)\bigl(\hat{b}+\hat{b}^\dagger\bigr)\,,
\end{multline}
and
\begin{multline}
\label{eq:EoMbSSPrecursor}
\dot{\hat{b}}=-(i\omega_\text{m}+\kappa_\text{m})\hat{b}-\sqrt{2\kappa_\text{m}}\,\hat{b}_\text{in}\\
-ig_0\beta\sqrt{\frac{L}{2c}}\Bigl[\bigl(\overline{a_\text{in}}^\ast\hat{a} + \overline{a}\,\hat{a}_\text{in}^\dagger\bigr)-\mathrm{H.c.}\Bigr]\,.
\end{multline}
\end{subequations}
In the adiabatic and small-coupling limit, \eref{eq:EoMaSSPrecursor} is formally solved by assuming that $\hat{b}$ varies slowly on the relevant timescale.
The solution to this equation is then substituted into \eref{eq:EoMbSSPrecursor} to yield a differential equation relating $\dot{\hat{b}}$ to the input fields and, in the rotating-wave approximation, exclusively to $\hat{b}$.
This equation of motion for $\hat{b}$ exhibits optically-induced damping of the micromirror and the optical spring effect; the micromirror oscillation frequency is $\omega_\text{m}+\omega_\text{o}$, where
\begin{align}
\omega_\text{o}=-g_0^2\lvert\beta&\rvert^2\frac{L}{2c}\lvert\overline{a_\text{in}}\rvert^2\frac{1}{\Delta^2+\kappa_\text{c}^2}\nonumber\\
\times\Biggl[&\frac{2\Delta\kappa_\text{c}^2+(\Delta+\omega_\text{m})(\kappa_\text{c}^2-\Delta^2)}{(\Delta+\omega_\text{m})^2+\kappa_\text{c}^2}\nonumber\\
&+\frac{2\Delta\kappa_\text{c}^2+(\Delta-\omega_\text{m})(\kappa_\text{c}^2-\Delta^2)}{(\Delta-\omega_\text{m})^2+\kappa_\text{c}^2}\Biggr]\,;
\end{align}
its damping constant is $\kappa_\text{m}+\kappa_\text{o}$, with
\begin{align}
\kappa_\text{o}=g_0^2\lvert\beta&\rvert^2\frac{L}{2c}\lvert\overline{a_\text{in}}\rvert^2\frac{\kappa_\text{c}}{\Delta^2+\kappa_\text{c}^2}\nonumber\\
\times\Biggl[&\frac{(2\Delta+\omega_\text{m})^2}{(\Delta+\omega_\text{m})^2+\kappa_\text{c}^2}-\frac{(2\Delta-\omega_\text{m})^2}{(\Delta-\omega_\text{m})^2+\kappa_\text{c}^2}\Biggr]\,.
\end{align}
After some algebra we can finally show that
\begin{align}
\label{eq:OccupationNumberFullSolution}
\expt{\hat{b}^\dagger\hat{b}}=&\ g_0^2\lvert\beta\rvert^2\frac{L}{2c}\lvert\overline{a_\text{in}}\rvert^2\frac{1}{\Delta^2+\kappa_\text{c}^2}\frac{1}{\kappa_\text{m}+\kappa_\text{o}}\nonumber\\
&\ \times\bigl[(\Delta-\omega_\text{m}-\omega_\text{o})^2+(\kappa_\text{c}+\kappa_\text{m}+\kappa_\text{o})^2\bigr]^{-1}\nonumber\\
&\ \times\bigl\{(\kappa_\text{m}+\kappa_\text{o})\bigl[\Delta^2+\kappa_\text{c}^2+\kappa_\text{c}(\kappa_\text{m}+\kappa_\text{o})\bigr]\nonumber\\
&\ \phantom{\times\bigl\{\ }+\kappa_\text{c}(2\Delta-\omega_\text{m}-\omega_\text{o})^2\bigr\}\nonumber\\
&+n_\text{th}\kappa_\text{m}/\bigl(\kappa_\text{m}+\kappa_\text{o}\bigr)\,.
\end{align}

\end{document}